\documentclass[11pt,twoside]{article}
\usepackage{asp2004}
\usepackage{psfig}
\usepackage{epsf}
\usepackage{graphics}
\usepackage{lscape}
\def\edcomment#1{\iffalse\marginpar{\raggedright\sl#1\/}\else\relax\fi}
\reversemarginpar

\begin{document}
\title{Supernova search at intermediate \boldmath $z$\\ II. Host galaxy morphology}
\author{J.~M\'endez~(1,2), P.~Ruiz-Lapuente~(1), G.~Altavilla~(1), A.~Balastegui~(1),
M.~Irwin~(3), K.~Schahmaneche~(4), C.~Balland~(5,6), R.~Pain~(4), N.~Walton~(3)}
\affil{1) Department of Astronomy, CER for Astrophysics, Particle
Physics and Cosmology, University of Barcelona, Diagonal 647, E--08028, 
Barcelona, Spain \\
2) Isaac Newton Group of Telescopes, 38700 Santa Cruz de La Palma, 
Islas Canarias, Spain \\
3) Institute of Astronomy, University of Cambridge, Madingley Road, 
Cambridge. CB3 0HA, United Kingdom\\
4) LPNHE-IN2P3-CNRS-Universit\'es Paris 6 et Paris 7, 4 place Jussieu, 
75252 Paris Cedex  05 France\\
5) Institut d'Astrophysique Spatiale, B\^{a}timent 121, Universit\'e
Paris 11, 91405 Orsay Cedex, France\\
6) Universit\'e Paris Sud, IAS-CNAS, B\^{a}timent 121,
Orsay Cedex, France}

\begin{abstract}
We discuss the host galaxy morphology of the 8 SNe
 discovered as a part
of the International Time Programme (ITP) project
``$\Omega$ and $\Lambda$ from Supernovae, and the Physics of Supernovae Explosions''
at the European Northern Observatory (ENO).
Identification of the SN host galaxy types was done 
exploiting both imaging and spectroscopy. A peculiar SNIa 
at z= 0.033 is found in a spiral galaxy, as most other SNeIa with z
between 0.1 and 0.4. A complete account of these studies will be given 
elsewhere.

\end{abstract}
\thispagestyle{plain}

\section{Galaxy morphology}
\begin{figure}
   \plottwo{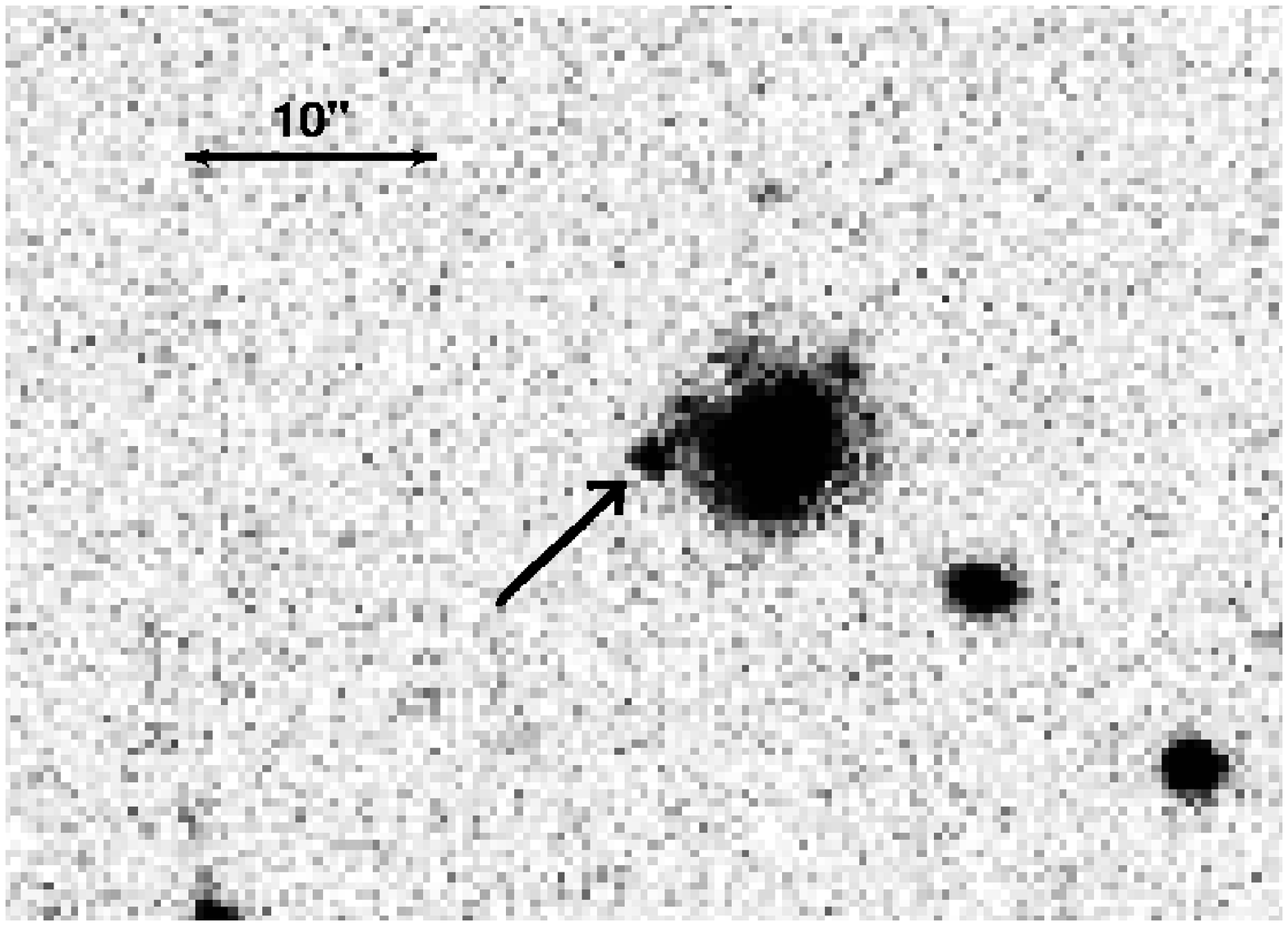}{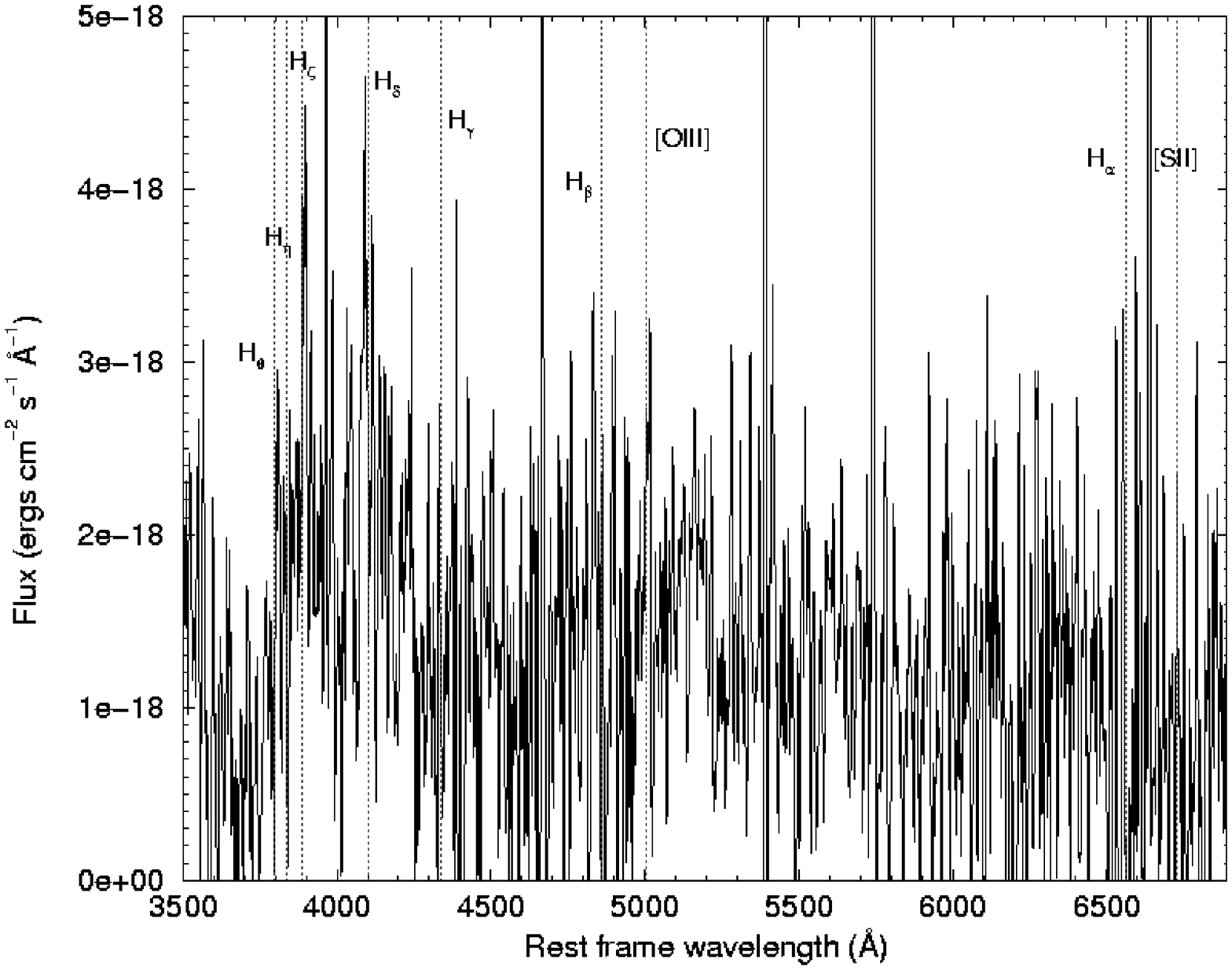}
   \caption{SN 2002lq host galaxy image (left panel) and spectrum (right panel).
    The SN position is marked by an arrow. The SN spectrum shows numerous  galaxy
    features (Balmer series lines are clearly visible). Both spectroscopy
    and imaging are consistent with a face-on spiral galaxy.}
   \label{SN2002lqimage} 
   \end{figure}

\begin{figure}
   \plottwo{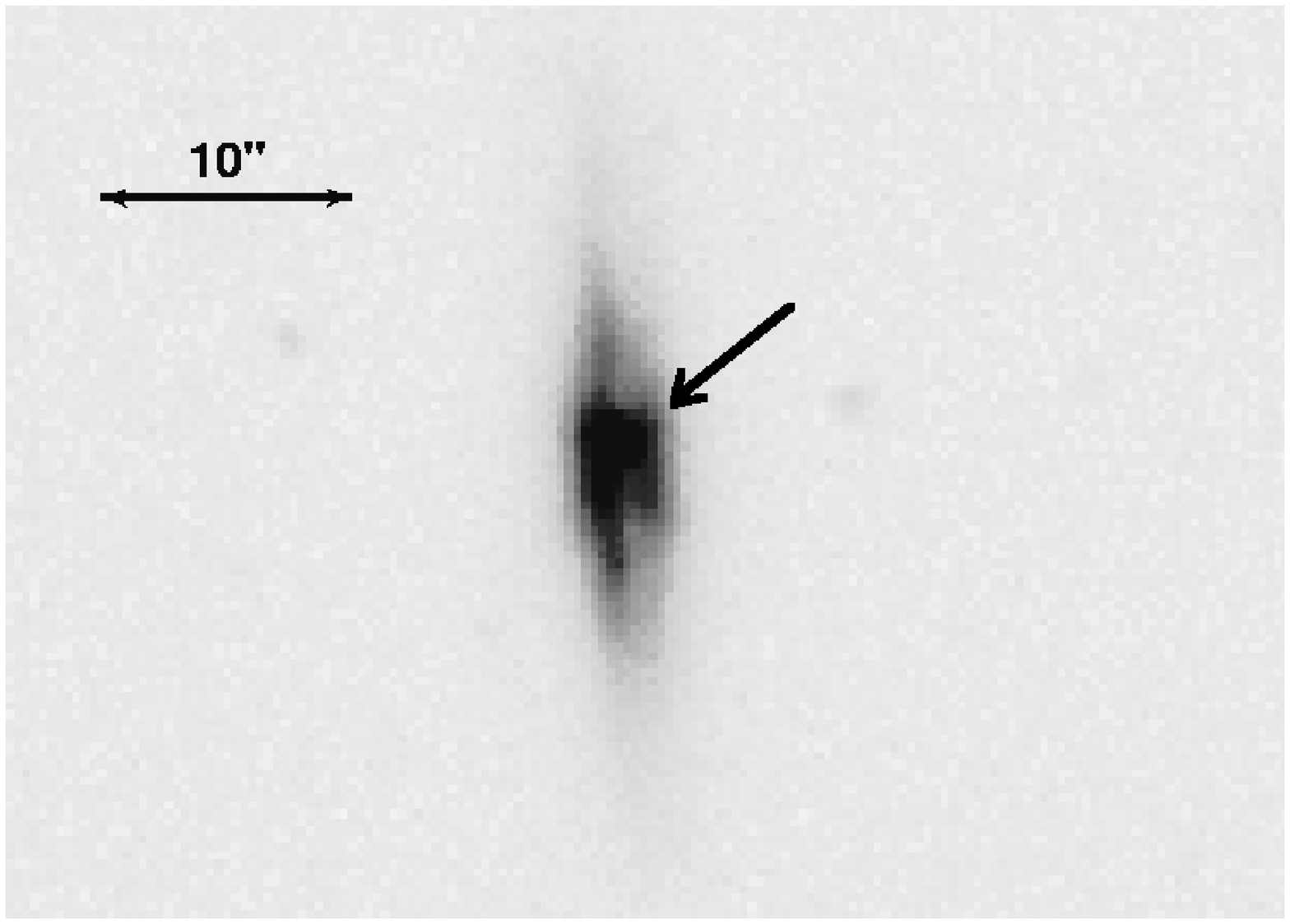}{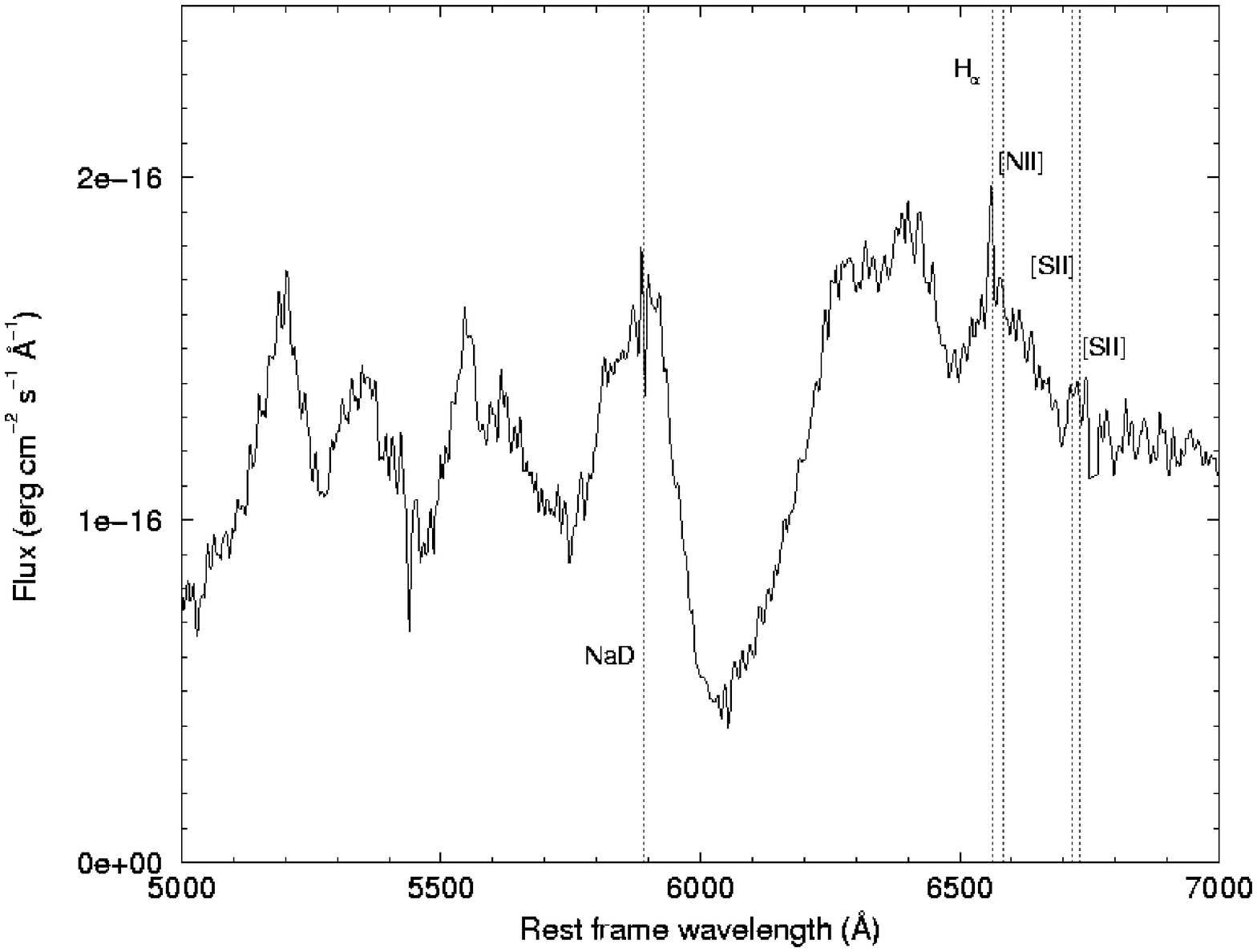}
   \caption{SN 2002lk host galaxy image (left panel) and spectrum (right panel).
    The SN position is marked by an arrow. The SN spectrum shows a few  galaxy
    features (due to the high brightness of the SN, the only galaxy lines
    clearly identified are $H_{\alpha}$, nitrogen and sulphur lines). 
    A deep sodium absorption is visible too. Spectroscopy and the absorption lane visible
    in the imaging suggest that this is an edge-on spiral galaxy.}
   \label{SN2002lkimage} 
   \end{figure}

Spectroscopic observations were carried out using 
 the Intermediate dispersion Spectrograph 
and Imaging System (ISIS) mounted at the 4.2m WHT.
Imaging was done using the WFC at the 2.5m INT, ALFOSC at the 2.5m
NOT and DOLORES at the 3.6m TNG.   
From the analysis of the galaxy spectra, though contaminated 
by the SN, we can discriminate
between spheroidal (elliptical and lenticular) and spiral galaxies.
In the analysed SN sample  only the morphologies of a few
host galaxies can be easily identified by visual inspection
(see Fig.\ref{SN2002lqimage}, \ref{SN2002lkimage}).
 The results are summarized  in Table \ref{tabla}\footnote{Host galaxy
 images of these SNeIa can be found at the
 web site of this project.}.
   \begin{table}
     \centering
      \caption[]{Host galaxy summary.}
         \label{tabla}
          \footnotesize

     $$ 
         \setlength\tabcolsep{2pt}
         \begin{tabular}{l  l  l  l   l l  l}
	    \hline
            \noalign{\smallskip}
 SN      &   $z$   & SN      &  Galaxy & \multicolumn{2}{c}{Offset}      & Identified lines\\
 name    &         & Type          &    Type     & \multicolumn{2}{c}{from nucleus} &  \\
            \hline
            \noalign{\smallskip}
2002li  &  0.329    & Ia     &  Spiral    & 0''.1 W& 0''.2 S  & H$_\alpha$, H$_\beta$, H$_\delta$, [SII], [NII], [OIII], [OII] \\
2002lj  &  0.180    & Ia     &  Spheroidal& \multicolumn{2}{c}{0''.2 W}          & - \\
2002lp  &  0.144    & Ia     &  Spiral    &0''.2 E & 2''.0 N  & H$_\alpha$, H$_\gamma$, H$_\theta$, [OII] \\
2002lq  &  0.269    & Ia     &  Spiral    &4''.5 E & 0''.7 S  & H$_\alpha$, H$_\beta$, H$_\gamma$, H$_\delta$, H$_\zeta$,  H$_\eta$, H$_\theta$, [SII], [OIII] \\
2002lr  &  0.255    & Ia     &  Spiral?   & \multicolumn{2}{c}{3''.0 S}         & H$_\beta$, [SII], [NII], [NeV]   \\
2002lk  &  0.033    & Ia pec.&  Spiral    & 0''.7 W& 0''.2 N & H$_\alpha$, [SII], [NII], NaD\\
\noalign{\smallskip}
2002ln  &  0.138    & II      & Spiral    & 1''.1 W& 8''.3 S & H$_\alpha$, H$_\beta$, H$_\zeta$,  H$_\theta$, [SII]  \\
2002lo  & 0.136     & II      & Spiral    & 0''.6 E& 1''.3 S & H$_\alpha$, H$_\beta$, H$_\theta$, [NII], [OIII], [OII]   \\
            \noalign{\smallskip}
            \hline
         \end{tabular}
     $$ 
\end{table}





\end{document}